# Open-Shell Molecules as Sensitive Probes of Spin–Orbit-Controlled Surface Electronic Structure

Solange Mariel Di Napoli[#1,2], Alexei Nefedov[#3], María Andrea Barral[#1,2], Gustavo E. Murgida[#1,2], Ana María Llois[1,2], Christof Wöll[*3], M. Verónica Ganduglia-Pirovano[4], and Verónica Laura Vildosola[*1,2]

[1] Departamento de Física de la Materia Condensada, Gerencia de Investigación y Aplicaciones, Comisión Nacional de Energía Atómica, Av. General Paz 1499, B1650KNA San Martín, Pcia. de Buenos Aires, Argentina.
[2] Instituto de Nanociencia y Nanotecnología CNEA-CONICET, Nodo Constituyentes, Av. General Paz 1499, B1650KNA San Martín, Pcia. de Buenos Aires, Argentina.
[3] Institute of Functional Interfaces (IFG), Karlsruhe Institute of Technology (KIT), Eggenstein-Leopoldshafen, Germany.
[4] Instituto de Catálisis y Petroleoquímica, Consejo Superior de Investigaciones Científicas (ICP-CSIC), 28049, Madrid, Spain.

# Authors contributed equally to this work.
* emails: christof.woell@kit.edu and veronicavildosola@integra.cnea.gob.ar

**Abstract:** Open-shell molecules are highly sensitive probes of correlated oxide surfaces because their partially occupied frontier orbitals strongly amplify subtle changes in adsorbate–substrate hybridization. Combining infrared reflection–absorption spectroscopy with hybrid density functional theory including non-collinear magnetism and spin–orbit coupling, we show that inclusion of spin–orbit coupling is essential to reproduce the vibrational spectrum of NO adsorbed on $UO_2$(111) by suppressing an otherwise artificial overhybridization between the NO frontier orbitals and uranium 5f states. While spin–orbit coupling determines the position of the N–O stretching band, dispersion-driven intermolecular interactions account for its asymmetric broadening at monolayer coverage. These results establish open-shell molecules as sensitive probes of spin–orbit-controlled surface electronic structure and identify NO/$UO_2$(111) as a stringent benchmark for electronic-structure methods describing open-shell molecule–surface interactions.

## Introduction

Chemical bonding on correlated oxide surfaces is governed by the interplay between charge transfer, orbital hybridization, magnetism, and relativistic effects. Although these ingredients are known to determine the electronic structure of actinide and transition-metal oxides, disentangling their individual roles in chemical bonding on surfaces remains a major challenge [1,2]. This problem is particularly acute for open-shell adsorbates, whose partially occupied frontier orbitals can strongly amplify subtle modifications of the substrate electronic structure into measurable changes in adsorption and vibrational properties. Beyond serving as exceptionally sensitive probes of surface electronic structure, open-shell molecules also interact with the local magnetic environment of the substrate, making them attractive building blocks for controlling molecular spin states and magnetic anisotropy and for applications in molecular spintronics and quantum information technologies [3-5]. Achieving a predictive understanding of these phenomena therefore requires electronic-structure methods capable of accurately describing electron correlation, non-collinear magnetism, and relativistic effects. Yet quantitative studies on structurally well-defined magnetic oxide surfaces remain scarce, underscoring the need for benchmark systems that enable direct comparison between experiment and theory and identify the level of electronic-structure treatment required for predictive descriptions of open-shell molecule–surface interactions.

Nitric oxide (NO) provides an ideal benchmark for this purpose. Unlike closed-shell molecules such as CO, NO possesses an unpaired electron occupying a partially filled antibonding π* orbital and a relatively small HOMO–LUMO gap [6]. Consequently, its adsorption behavior is exceptionally sensitive to subtle variations in adsorbate–substrate hybridization and charge redistribution. These changes are directly reflected in the N–O stretching frequency, making vibrational spectroscopy an unusually sensitive probe of the local electronic structure. At the same time, the open-shell character of NO poses a stringent challenge for electronic-structure methods, since semilocal density functionals substantially underestimate its frontier-orbital gap and consequently tend to overestimate molecule–surface hybridization and chemical bonding. The pronounced sensitivity of the N–O stretching vibration makes infrared reflection–absorption spectroscopy (IRRAS) on oxide single crystals an exceptionally powerful experimental probe. Although experimentally demanding owing to the weak infrared reflectivity of dielectric substrates compared with metals, single-crystal IRRAS directly connects adsorption geometry, local electronic structure, and vibrational fingerprints, making it uniquely well suited for quantitative comparisons between experiment and theory of open-shell molecule–surface interactions on correlated oxide surfaces [7].

Uranium dioxide constitutes a unique substrate on which to exploit this sensitivity. As the prototypical correlated actinide oxide, $UO_2$ combines strongly correlated 5f electrons, non-collinear magnetic order (NCL), and pronounced spin–orbit coupling (SOC), all of which are essential for reproducing its ground-state electronic structure [8-11]. Because the frontier orbitals of NO interact directly with the uranium 5f manifold, the NO/$UO_2$ (111) interface provides a well-defined model system for disentangling the respective roles of electron correlation, non-collinear magnetism, and relativistic effects in molecule–surface interactions.

Here, we combine infrared reflection–absorption spectroscopy (IRRAS) with hybrid density functional theory including non-collinear magnetism, spin–orbit coupling, and dispersion interactions to investigate NO adsorption on $UO_2$(111). We demonstrate that only the combined treatment of hybrid exchange, non-collinear magnetism, and spin–orbit coupling accurately reproduces the experimental vibrational spectrum because spin–orbit coupling suppresses an otherwise artificial overhybridization between the NO frontier orbitals and the uranium 5f states. Comparison with the closed-shell probe molecule CO reveals that the open-shell electronic structure of NO amplifies subtle spin–orbit-induced modifications of the substrate into large and directly measurable changes in adsorption geometry and vibrational properties. At monolayer coverage, configurational diversity and dispersion-driven lateral interactions generate a manifold of collective vibrational modes that account for the broad asymmetric infrared spectrum. More generally, our results establish open-shell molecules as highly sensitive probes of spin-orbit-controlled surface electronic structure and provide a stringent benchmark for electronic-structure methods describing open-shell molecule–surface interactions on correlated oxides.

**Experimental**

Infrared reflection–absorption spectroscopy (IRRAS) spectra recorded after NO adsorption on $UO_2$(111) at 113 K are shown in Fig. 1b (experimental details are provided in the Supporting Information, Section 1 [19]). The spectra are dominated by a strong absorption band centered at approximately 1860 $cm^{-1}$, accompanied by a weaker shoulder near 1825 $cm^{-1}$. The position of the main band is close to that reported for molecular NO adsorbed on $TiO_2$(110), consistent with weak molecular chemisorption [12]. In contrast, NO adsorption on many transition-metal surfaces is accompanied by substantially stronger adsorbate–surface interactions, resulting in N–O stretching frequencies typically below 1800 $cm^{-1}$. Unlike $TiO_2$(110), however, no vibrational features associated with ON-NO dimers (1750 $cm^{-1}$) or $N_2O$ (2243 $cm^{-1}$) are observed, indicating that NO remains molecular on $UO_2$(111) under the present experimental conditions. A possible interpretation of the lower-frequency shoulder is adsorption at oxygen vacancies, analogous to the 10–20 $cm^{-1}$ vacancy-induced red shifts reported for CO on $CeO_2$(111) [13]. However, if the shoulder were due to NO adsorbed at oxygen vacancies on $UO_2$(111), the expected charge transfer from the reduced substrate would lead to the formation of $NO^-$ (or $NO^{\delta-}$) species, accompanied by substantially larger red shifts, typically exceeding 100 $cm^{-1}$. A defect-related origin of the shoulder would also be inconsistent with previous IRRAS studies of CO adsorption on the same $UO_2$(111) surface [14], which revealed only a single narrow vibrational band, providing no evidence for a significant concentration of defect sites. These observations therefore strongly suggest that the shoulder at 1825 $cm^{-1}$ is not associated with surface defects but is instead an intrinsic feature of NO adsorption on the stoichiometric $UO_2$(111) surface. Elucidating the microscopic origin of this feature, and the contrasting vibrational response of NO and CO on $UO_2$(111), requires a detailed theoretical analysis.

## Theoretical results

To elucidate the microscopic origin of the experimental spectra, we performed first-principles calculations (computational details are provided in the Supporting Information, Section 2 [19]). Hybrid HSE06 [15-16] calculations including Grimme D3 dispersion corrections [17-18], non-collinear (NCL) magnetism, and spin–orbit coupling (SOC) reproduce the experimental IRRAS spectrum remarkably well (Fig. 1). Hereafter, PBE+U (CL), HSE06 (CL), HSE06 (NCL), and HSE06+SOC (NCL) denote PBE+U and HSE06 calculations performed within the collinear formalism, HSE06 calculations performed within the non-collinear formalism, and HSE06 calculations performed within the non-collinear formalism including spin–orbit coupling, respectively. Fig. 1a displays the calculated N–O stretching frequencies and corresponding IR intensities for the different 1 ML NO adsorption configurations considered (see Fig. S4). Despite the variety of adsorption configurations and surface periodicities, the calculated spectra consistently reveal two characteristic groups of vibrational frequencies associated with two underlying local adsorption motifs, namely top-like and bridge-like. The larger surface periodicities introduce additional structural distortions and collective vibrational modes while retaining the characteristic frequency groups associated with the two local adsorption motifs. The representative (1×1) top-like and bridge-like structures shown in Fig. 1c, d exhibit stretching frequencies of 1862 and 1856 $cm^{-1}$ (Table 1), in excellent agreement with the experimental band centered at 1860 $cm^{-1}$.

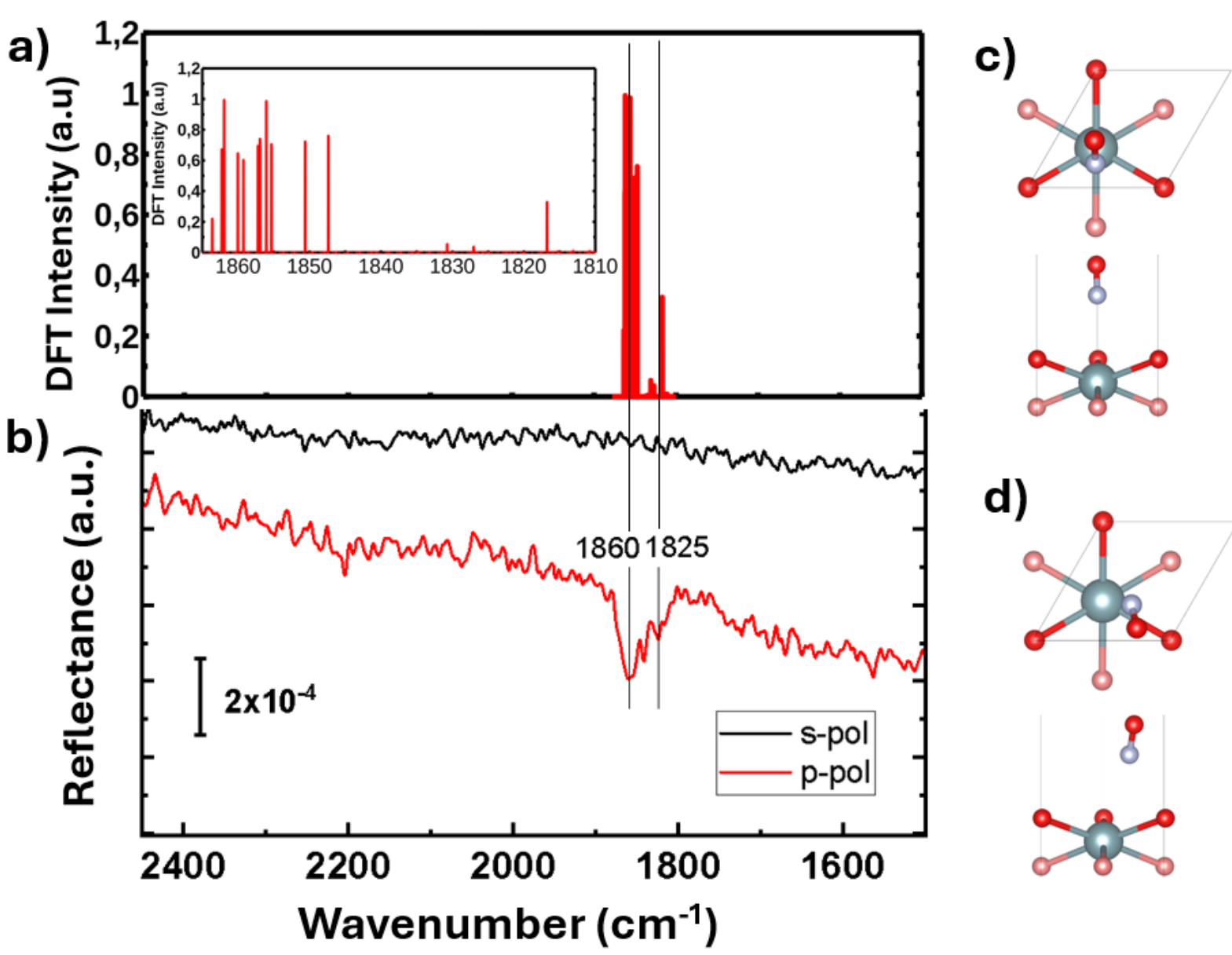

**Figure 1:** (a) Calculated N–O stretching frequencies and corresponding IR intensities per NO molecule obtained using HSE06+SOC (NCL) for 1 ML NO overlayers on $UO_2$ (111) with different surface periodicities (1×1, 2×1, 3×1, and 2×2). Calculated IR intensities are first normalized per NO molecule and then scaled so that the highest calculated intensity equals unity. (b) IRRAS spectra recorded after saturation NO adsorption on $UO_2$ (111) at 113 K using *s*- and *p*-polarized light. Top and side views of the optimized (c) top-like and (d) bridge-like adsorption geometries for the (1×1) structure.

**Table 1.** Scaled N–O stretching frequencies ($cm^{-1}$) calculated using HSE06+SOC (NCL) for 1 ML NO overlayers on $UO_2$ (111) using different surface periodicities (1×1, 2×1, 3×1, and 2×2). The table lists all IR-active N–O stretching modes obtained for each surface periodicity. Superscript numbers identify the adsorption configuration (see Fig. S4), whereas the attached superscript letters indicate the calculated IR intensity ($^{s}$, strong; $^{m}$, medium; $^{w}$, weak). The columns $v_1$–$v_4$ correspond to the collective N–O stretching modes of the NO overlayer, listed in decreasing frequency. Modes with zero IR intensity are omitted.

| Surface periodicity | $v_1$ | $v_2$ | $v_3$ | $v_4$ |
|---|---|---|---|---|
| **(1×1)** | $1862^{1s}$ | – | – | – |
| | $1856^{2s}$ | | | |
| **(2×1)** | $1860^{3s}$ | – | – | – |
| | $1855^{4s}$ | $1825^{4w}$ | | |
| | $1862^{5s}$ | $1831^{5s}$ | | |
| | $1859^{6s}$ | $1827^{6m}$ | | |
| **(3×1)** | $1862^{7s}$ | – | – | – |
| | $1857^{8s}$ | $1835^{8w}$ | $1831^{w}$ | |
| **(2×2)** | $1864^{9s}$ | $1817^{9s}$ | $1757^{9m}$ | $1473^{9s}$ |
| | $1857^{10s}$ | – | – | – |
| | $1851^{11s}$ | – | $1815^{11w}$ | $1813^{11w}$ |
| | $1847^{12s}$ | $1821^{12w}$ | – | $1811^{12w}$ |

Although the experimental IRRAS spectrum is reproduced remarkably well by the complete HSE06+SOC (NCL) description, this agreement relies critically on an accurate treatment of the electronic structure of both the $UO_2$ substrate and the NO adsorbate. To identify the essential components of the theoretical description, we systematically examined the influence of the exchange-correlation functional, the magnetic formalism, and spin–orbit coupling using the representative top-like (1×1) adsorption motif (Table 2), with additional comparisons presented in the Supporting Information, Section 4[19]. As summarized in Table 2, the N–O stretching frequency exhibits an exceptional sensitivity to the level of theory. While the complete HSE06+SOC (NCL) treatment reproduces the experimental frequency within only 2 $cm^{-1}$, omitting SOC lowers the calculated frequency by 74 $cm^{-1}$, and a collinear (CL) HSE06 description results in an even larger deviation of 148 $cm^{-1}$. These changes are accompanied by a marked shortening of the N–U distance, and elongation of the N–O bond, and stronger adsorption energies, indicating a substantial overestimation of the adsorbate–substrate interaction and a concomitant weakening of the N–O bond.

**Table 2.** Effect of the electronic-structure description on the structural, energetic, and vibrational properties of NO adsorbed in the representative top-like (1×1) motif on $UO_2$(111). Results are shown for collinear PBE+U (CL), collinear HSE06 (CL), non-collinear HSE06 (NCL) without spin–orbit coupling, and non-collinear HSE06 including spin–orbit coupling (HSE06+SOC (NCL)). The corresponding bridge-like HSE06+SOC (NCL) result is included for comparison. Listed are the N–O bond length *d*(N–O), its variation with respect to the gas-phase molecule Δ*d*(N–O), the N–U and N–$O_s$ (surface oxygen) bond distances, the adsorption energy $E_{ads}$, the scaled N–O stretching frequency $\nu$, and the corresponding frequency shift $\Delta\nu$ with respect to the gas phase. See Table S1 for the corresponding scaling factors.

| Motif | Method | *d*(N–O)[pm] | Δ*d*(N–O)[pm] | *d*(N–U) [Å] | *d*(N–$O_s$) [Å] | $E_{ads}$ [eV] | $\nu$ [$cm^{-1}$] | $\Delta\nu$ [$cm^{-1}$] |
|---|---|---|---|---|---|---|---|---|
| Top-like | PBE+U (CL) | 117.6 | +0.54 | 2.67 | 2.66 | -0.44 | 1818 | -58 |
| | HSE06 (CL) | 117.8 | +2.30 | 2.23 | 2.84 | -0.47 | 1714 | -164 |
| | HSE06 (NCL) | 116.5 | +1.00 | 2.39 | 2.85 | -0.39 | 1788 | -87 |
| | HSE06+ SOC (NCL) | 115.5 | -0.02 | 2.81 | 2.90 | -0.35 | 1862 | -14 |
| Bridge-like | HSE06+ SOC (NCL) | 115.6 | 0.08 | 2.83 | 2.51 | -0.37 | 1856 | -20 |

The pronounced dependence of the adsorption properties on the level of theory reflects the distinct physical roles of the different components of the electronic structure description. The commonly employed collinear PBE+U approach yields a stretching frequency of 1818 $cm^{-1}$, substantially closer to experiment than collinear HSE06. This apparent improvement, however, should not be interpreted as a superior description of the adsorption process, but most likely reflects a fortuitous cancellation of errors. In fact, within PBE+U approach the NO molecule is described at the PBE level, which already severely underestimates the gas-phase HOMO-LUMO gap (see section 4 of Supporting Information [19]). The failure of HSE06 (CL) does not originate from the hybrid functional itself, but from the absence of the magnetic and relativistic ingredients required for an adequate description of the uranium 5*f* electronic structure. As evidenced by the systematic structural, energetic, and vibrational trends summarized in Table 2, the HSE06 hybrid functional reduces the self-interaction error, providing a more realistic description of both the localized U 5*f* states and the electronic structure of NO. The non-collinear formalism removes the artificial constraint that all uranium magnetic moments remain aligned along a common quantization axis, allowing the complex magnetic order characteristic of $UO_2$ to develop. Finally, spin–orbit coupling, a fundamental relativistic interaction in actinide compounds, accounts for the coupling between the spin and orbital angular momenta of the uranium 5*f* electrons, further reshaping the 5*f* electronic structure and its hybridization with the adsorbed NO molecule. The evolution of the electronic and magnetic structure across these theoretical levels of theory is documented in Table S2 of the Supporting Information. [19]

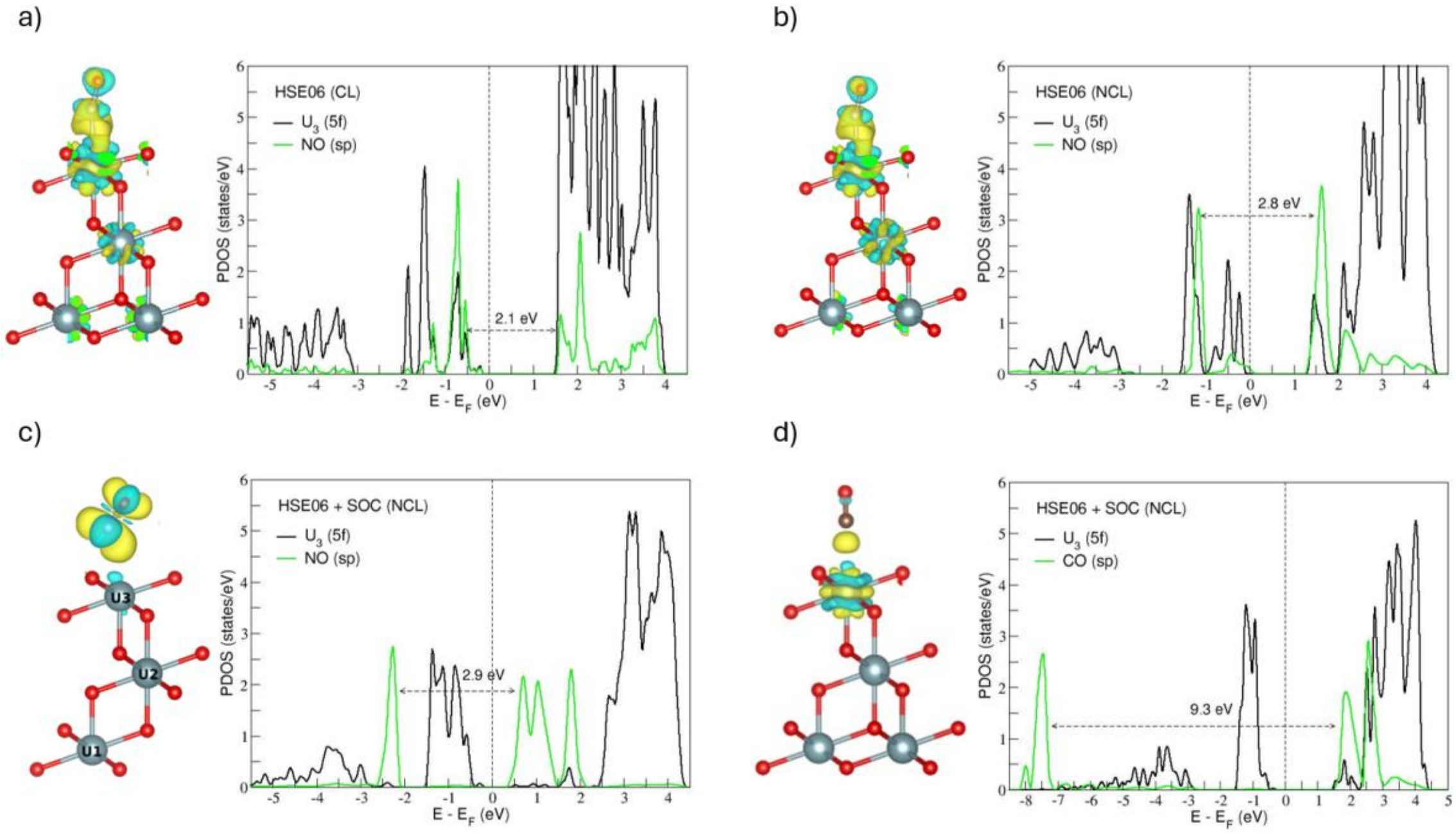

**Figure 2.** Optimized adsorption geometries with the corresponding charge-density difference (left) and the Projected Density of States (PDOS, right). Panels (a-c) show NO adsorbed on $UO_2$(111) calculated using (a) HSE06 (CL), (b) HSE06 (NCL), and (c) HSE06+SOC (NCL). Panel (d) shows CO adsorbed on $UO_2$(111) calculated using HSE06+SOC (NCL). Uranium, nitrogen, carbon, and oxygen atoms are shown in dark grey, light grey, brown, and red, respectively. The PDOS correspond to the 5*f* states to the surface uranium atom directly bonded to the adsorbate ($U_3$) and to the molecular orbitals of the adsorbate. The charge density difference is shown for an isosurface value of ±0.0034 eÅ$^{-3}$ in all panels; yellow and light blue isosurfaces denote charge accumulation and depletion, respectively.

The microscopic origin of these trends is revealed by the projected densities of states shown in Fig. 2. The corresponding charge-density difference and PDOS for PBE+U (CL) are provided in Fig. S5 of the Supporting Information. [19] In the absence of spin–orbit coupling, i.e., for HSE06 (CL) and HSE06 (NCL) (Fig. 2a,b), the NO-derived states of the adsorbed molecule exhibit substantially stronger overlap with the U 5*f* manifold, indicating an artificial enhancement of the adsorbate–substrate hybridization. Upon inclusion of spin–orbit coupling (HSE06+SOC (NCL), Fig. 2c), the coupled NO/$UO_2$ electronic structure is reorganized: the overlap between NO-derived states and the U 5*f* manifold is markedly reduced, leading to weaker adsorbate–substrate covalent interaction, a longer N–U bond, recovery of the intrinsic N–O bond strength, and a corresponding improvement in the calculated stretching frequency.

The exceptional sensitivity of the calculated adsorption properties becomes even more apparent when compared with our previous study of CO adsorption on $UO_2$(111) [14]. Whereas inclusion of spin–orbit coupling modified the calculated C–O stretching frequency by only about 3 cm$^{-1}$, the corresponding shift for NO reaches 74 cm$^{-1}$. This striking difference originates from the fundamentally different electronic structures of the two molecules. Unlike closed-shell CO, NO possesses a singly occupied π* orbital and a considerably smaller HOMO–LUMO gap, making its frontier electronic states much more susceptible to subtle changes in the electronic structure of the substrate. Consequently, the SOC-induced modification of the uranium 5*f* states leads to much larger changes in adsorbate–substrate hybridization, adsorption geometry, and vibrational frequency than in the case of CO.

The different bonding mechanisms are further illustrated by the charge-density differences shown in the left panels of Fig. 2c,d. Although both molecules adsorb molecularly through relatively weak chemisorption, the balance between charge donation and back-donation differs substantially. For CO, electron donation from the molecule to the substrate dominates, resulting in a slight strengthening of the internal C–O bond and the experimentally observed blue shift of the stretching frequency. In contrast, adsorption of NO is characterized by a net charge transfer from the substrate to the molecule, consistent with the weakening of the N–O bond and the corresponding red shift of the stretching frequency.

Unlike CO, however, the vibrational response of adsorbed NO cannot be understood solely in terms of the internal bond strength. Despite the pronounced red shift, the N–O bond length changes only marginally. Analysis of the vibrational eigenvectors reveals that the N–O stretching mode remains predominantly localized on the molecule but exhibits approximately an order-of-magnitude larger participation of the exposed uranium atom and neighboring surface oxygen atoms than in the case of CO. Consequently, the observed stretching frequency is determined not only by the intrinsic N–O bond strength but also by the dynamical coupling between the molecule and the surface atoms, making the vibrational response considerably more sensitive to the adsorption geometry and electronic structure of the interface. A simple mechanical model based on coupled restoring forces provides an intuitive interpretation of this dynamical coupling and is discussed in Section 5 of the Supporting Information. [19]

These results demonstrate that NO is an exceptionally sensitive probe of the  the spin–orbit-controlled structure of $UO_2$. Its open-shell electronic structure amplifies subtle SOC-induced modifications of the uranium 5*f* states into large and directly measurable changes in adsorption and vibrational properties, whereas the corresponding effects remain comparatively small for conventional closed-shell probe molecules such as CO.

The robustness of these conclusions with respect to the magnetic structure of $UO_2$ was further examined by considering several non-collinear magnetic configurations of the slab, as well as different relative orientations of the NO magnetic moment with respect to the surface uranium moments (see Sections 4, 6, and 7 of Supporting Information).[19] Although these configurations produce variations in the adsorption energies of up to ~70 meV (Fig. S7) and magnetic coupling energies below 20 meV (Fig. S8), the calculated N–O stretching frequencies remain essentially unchanged. These results demonstrate that the pronounced SOC dependence identified above is an intrinsic consequence of the electronic structure of the NO/$UO_2$ interface rather than an artifact of a particular magnetic configuration, supporting the applicability of the present 0 K calculations to the experimental conditions (113 K), where $UO_2$ is paramagnetic.

The calculations presented above explain the position of the main absorption band but not its pronounced asymmetry and the shoulder observed experimentally around 1825 $cm^{-1}$.

The representative (1×1) calculations reproduce the frequency of the dominant experimental band. However, they describe only the in-phase collective stretching mode of a perfectly ordered NO overlayer. The experimental spectrum is considerably broader and asymmetric, indicating that additional vibrational contributions must be present. As discussed above, a defect-related origin of the shoulder was ruled out. To investigate this effect, we considered larger surface periodicities and multiple 1 ML adsorption arrangements (Fig. S4), thereby allowing intermolecular coupling and configurational diversity within the adlayer.

For each adsorption arrangement, the lateral interaction between neighboring NO molecules gives rise to several collective N–O stretching modes with non-zero IR intensity (Table 1). Although the highest-frequency mode remains close to the experimental maximum, additional IR-active modes appear at lower frequencies. The resulting manifold of closely spaced vibrational modes, arising from intermolecular coupling within the NO adlayer together with the adsorbate–substrate dynamical coupling discussed above, naturally accounts for both the broad asymmetric line shape and the shoulder around 1825 $cm^{-1}$ observed experimentally (Fig. 1). Although additional adsorption arrangements may exist, the configurations considered here already capture the essential effects of intermolecular coupling on the infrared response.

The emergence of these collective vibrational modes implies that lateral interactions between neighboring NO molecules are sufficiently strong to influence the infrared response of the saturated adlayer. To disentangle the intrinsic intermolecular interaction from substrate-mediated effects, we analyzed the interaction between isolated pairs of NO molecules as a function of their separation (see Section 8 of Supporting Information for computational details [19]). Although performed in vacuum, these calculations isolate the direct intermolecular interaction, allowing its contribution to be separated from the configurational and substrate-mediated effects present in the adsorbed overlayer. Fig. 3 shows the interaction energy between two NO molecules as a function of their intermolecular separation in vacuum. As expected, the interaction is attractive at intermediate intermolecular separations. At shorter distances, overlap of the molecular charge densities leads to the onset of short-range Pauli repulsion, making the interaction progressively less attractive.

The same analysis performed for CO reveals substantially weaker intermolecular interactions. At the nearest-neighbor U–U distance on $UO_2(111)$ (3.85 Å), the interaction between NO molecules remains significantly more attractive than for CO, indicating a much stronger propensity for lateral coupling between neighboring NO molecules.

To identify the origin of the attractive interaction between NO molecules, additional calculations were performed without dispersion corrections, revealing that van der Waals interactions provide the dominant attractive contribution at intermediate intermolecular separations. The NO curve calculated without dispersion corrections lies close to zero at the nearest-neighbor U–U distance, indicating that the residual attractive interaction is very small (~2 meV). Additional calculations comparing parallel and antiparallel molecular spin orientations give energy differences below 0.3 meV, showing that magnetic interactions contribute negligibly to the intermolecular attraction, even at 0 K.

Upon adsorption, these attractive dispersion forces compete with the electrostatic repulsion arising from the partial charge transfer associated with weak chemisorption. In fact, omitting van der Waals corrections significantly weakens the lateral interactions (see Supporting Information, Section 4 [19]), causing the oscillator strength to become concentrated in the collective in-phase stretching mode while the remaining collective modes lose intensity. The interplay between the attractive dispersion forces and the electrostatic repulsion provides a microscopic rationale for the strong lateral interactions within the NO adlayer. Together with configurational diversity and the dynamical interaction with the substrate, these lateral interactions give rise to the manifold of closely spaced collective vibrational modes that accounts for the experimentally observed broad asymmetric infrared band and the shoulder at 1825 $cm^{-1}$.

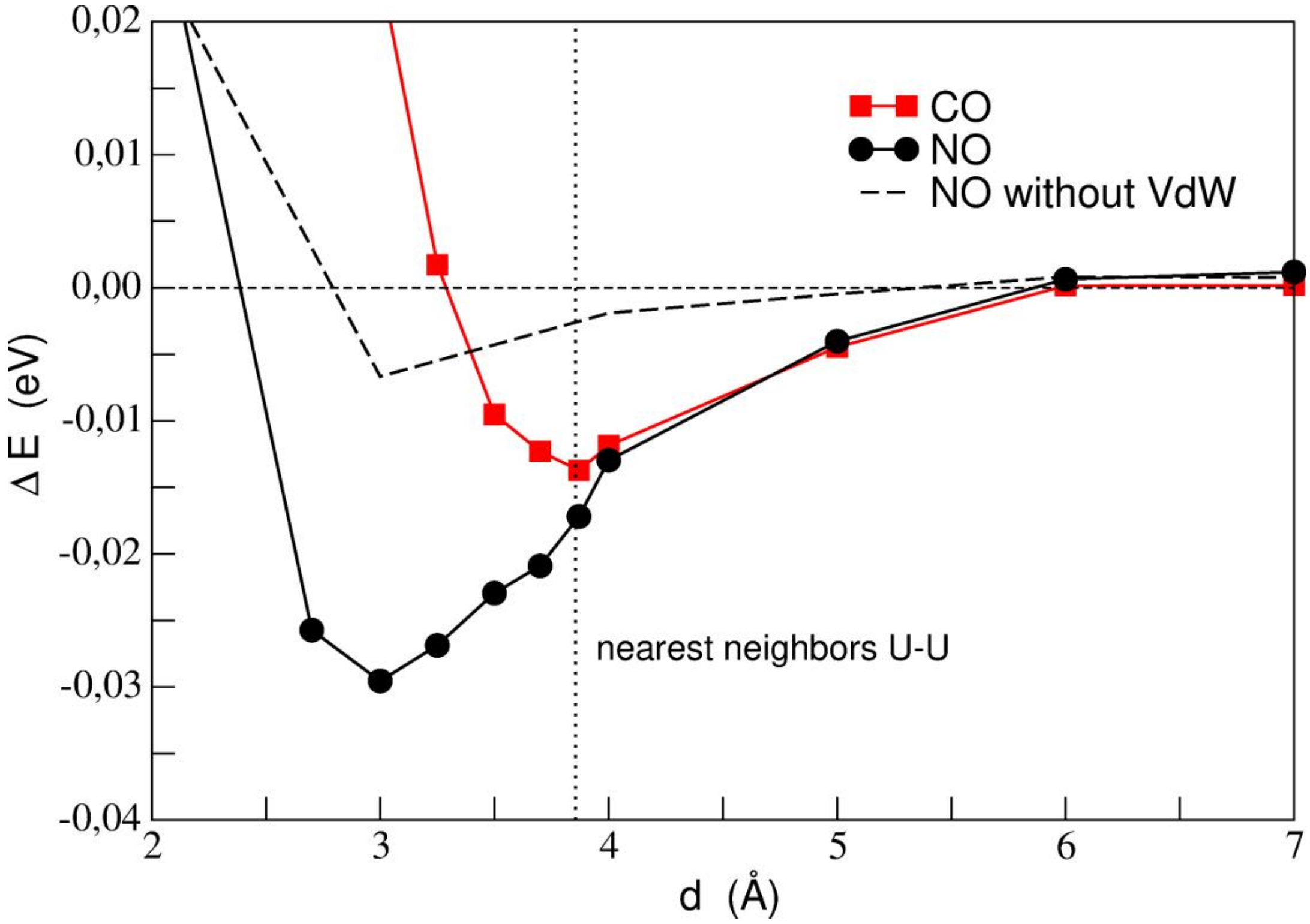

**Figure 3.** Interaction energy of two NO or CO molecules in vacuum as a function of the intermolecular separation calculated in a 20 x 11 x 12 $Å^3$ supercell. The interaction energies are referenced to the largest separation considered (d= 10 Å). For NO, additional calculations were performed without dispersion (vdW) corrections. The fine dotted line indicates the nearest-neighbor U–U distance on $UO_2$(111) (d=3.85 Å).

In summary, we show that a quantitatively reliable description of NO adsorption on $UO_2$(111) requires an accurate treatment of both the electronic structure of the adsorbate–substrate interface and the lateral interactions within the adsorbed layer. Hybrid exchange, non-collinear magnetism, and spin–orbit coupling are essential to reproduce the adsorption geometry and vibrational frequency of isolated NO, whereas dispersion interactions govern the lateral coupling responsible for the broad asymmetric infrared band at monolayer coverage.

Unlike closed-shell CO, the open-shell electronic structure of NO amplifies subtle spin–orbit-induced modifications of the uranium 5*f* states into large and directly measurable changes in adsorption geometry and vibrational properties. Comparison with CO further reveals fundamentally different charge-transfer mechanisms, with CO acting predominantly as an electron donor and NO as an electron acceptor from the surface. More generally, our results establish open-shell molecules as highly sensitive probes of spin–orbit-controlled surface electronic structure and highlight the need to account simultaneously for relativistic electronic structure and intermolecular interactions when interpreting vibrational spectra of correlated oxide surfaces.

**Acknowledgements**

We thank Prof. Hicham Idriss for providing the $UO_2$ single crystal used in the experiments and for his support during the initial stages of this project. MAB, SDN, GM, AMLL, and VV thank the financial support from by Agencia Nacional de Promoción Científica y Tecnológica (ANPCyT) through the project PICT 2019/02129 and from Consejo Nacional de Investigaciones Científicas y Técnicas (CONICET) for the project PIP 2023-2025 11220220100330CO. MVGP gratefully acknowledges funding from projects PID2021-128915NB-I00 and PID2024-162810NB-I00 financed by Ministerio de Ciencia, Innovacion y Universidades, and ERDF, EU. Computational resources were provided by the Red Española de Supercomputación (RES) at MareNostrum 5 (BSC, Barcelona).

**Data availability**

The data that support the findings of this study are available in the supporting information of this article. The DFT data that support the findings of this study are available in Materials Cloud {https://www.materialscloud.org/ home} with the identifier DOI: ..... The data are also available from the authors upon reasonable request.

## Supporting Information

**Open-Shell Molecules as Sensitive Probes of Spin–Orbit-Controlled Surface Electronic Structure**

Solange Mariel Di Napoli[#1,2], Alexei Nefedov [#3], María Andrea Barral[#1,2], Gustavo E. Murgida[#1,2], Ana María Llois [1,2], Christof Wöll[*3], M. Verónica Ganduglia-Pirovano[4], and Verónica Laura Vildosola[* 1,2]

[1] Departamento de Física de la Materia Condensada, Gerencia de Investigación y Aplicaciones, Comisión Nacional de Energía Atómica, Av. General Paz 1499, B1650KNA San Martín, Pcia. de Buenos Aires, Argentina.
[2] Instituto de Nanociencia y Nanotecnología CNEA-CONICET, Nodo Constituyentes, Av. General Paz 1499, B1650KNA San Martín, Pcia. de Buenos Aires, Argentina.
[3] Institute of Functional Interfaces (IFG), Karlsruhe Institute of Technology (KIT), Eggenstein-Leopoldshafen, Germany.
[4] Instituto de Catálisis y Petroleoquímica, Consejo Superior de Investigaciones Científicas (ICP-CSIC), 28049, Madrid, Spain
# Authors contributed equally to this work.
* emails: christof.woell@kit.edu and veronicavildosola@integra.cnea.gob.ar

### 1. Experimental Methods

The IRRAS measurements were performed in the advanced ultrahigh-vacuum (UHV) apparatus "THEO", which integrates a state-of-the-art Fourier-transform infrared (FTIR) spectrometer (Bruker Vertex 80v) with a UHV system (Prevac, Poland) comprising multiple interconnected chambers equipped with X-ray photoelectron spectroscopy (XPS), low-energy electron diffraction (LEED), and sample-preparation facilities. The apparatus is designed for high-sensitivity infrared spectroscopy and enables both grazing-incidence IRRAS measurements on oxide single crystals and UHV-FTIR transmission measurements on oxide powders. [20, 7, 21-22]

A natural $UO_2$ single crystal (approximately 8 × 3 × 1 $mm^3$) with a surface polished parallel to the (111) crystallographic plane was used. Further details about the sample preparation are described in Ref. 14. Prior to the IRRAS measurements, the $UO_2$(111) surface was additionally cleaned by repeated cycles of sputtering (1.5 keV, 6.5 mA, $p_{Ar}$=5 × $10^{-6}$ mbar, T=300 K, 30 min), followed by annealing in $O_2$ at 900 K for 45 min. Surface cleanliness and crystallographic order were verified by LEED (Fig. S1) and XPS (Fig. S2).

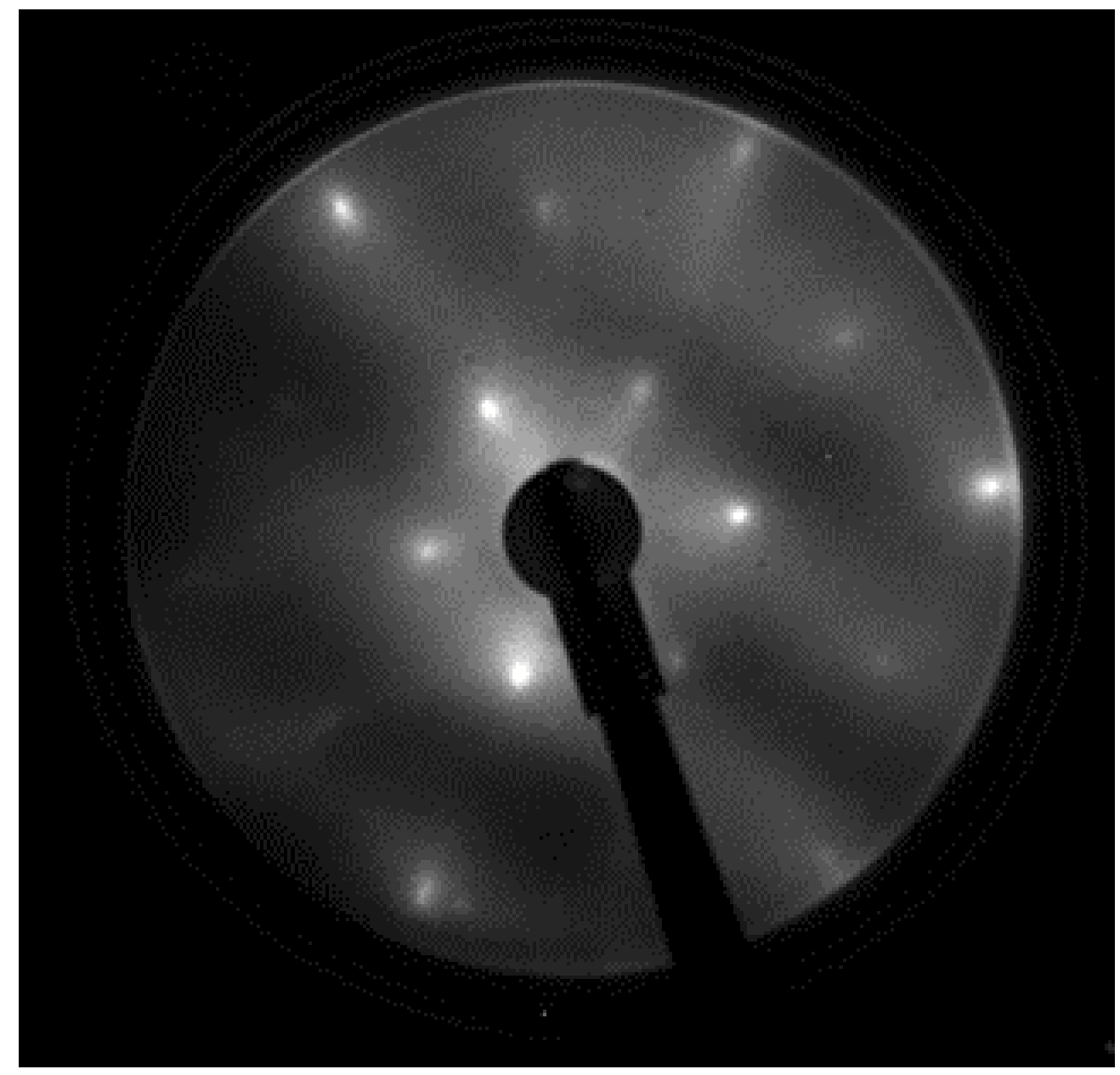

**Figure S1:** LEED image (E=127 eV) of the clean $UO_2$(111) surface. The sharp hexagonal diffraction pattern is consistent with previous observations for $UO_2$(111) [14] and confirms the (111) termination.

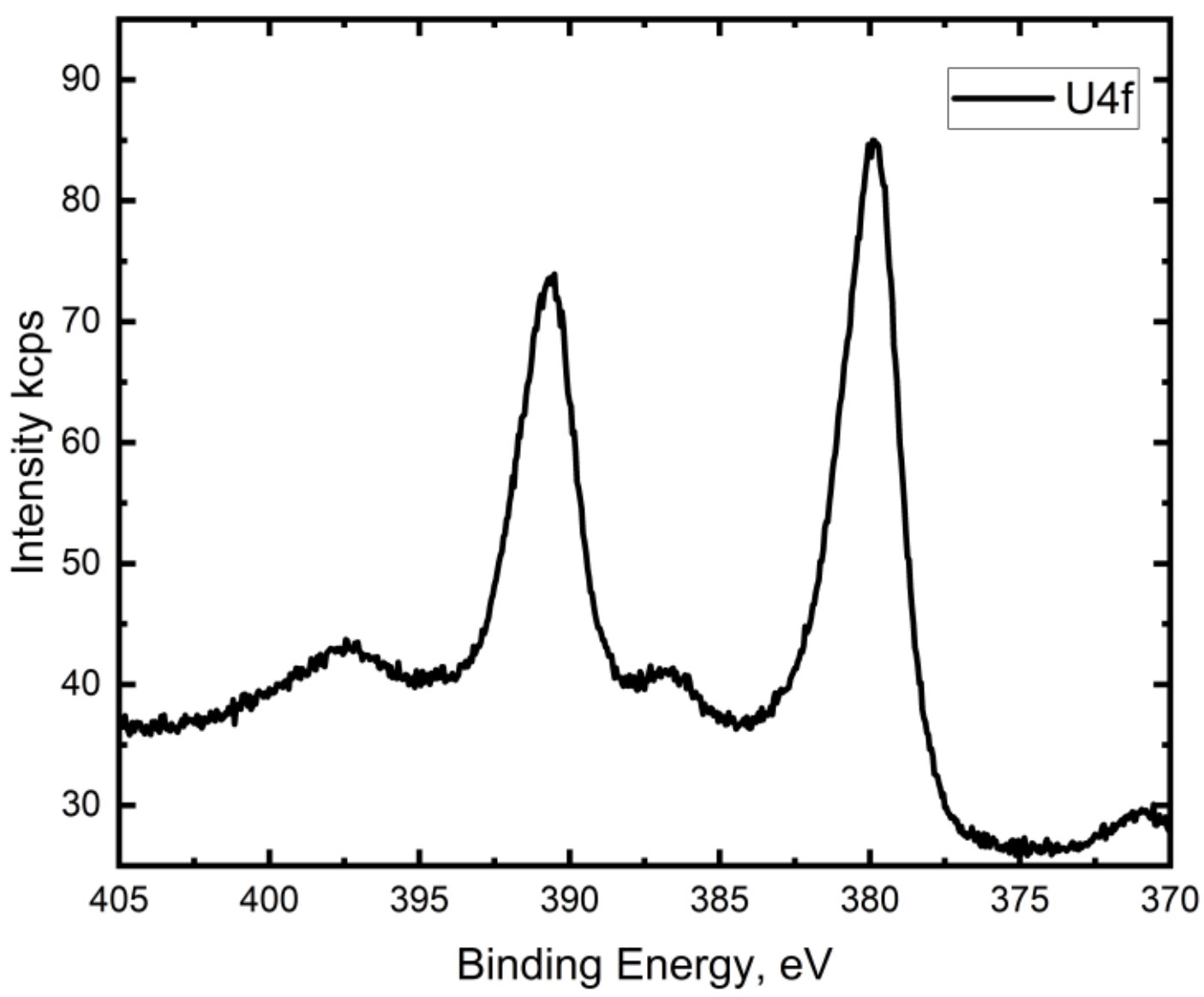

**Figure S2:** U 4f photoelectron spectrum using an Mg Kα source (*hv* =1253 eV). The spectrum is consistent with previous work [14] and confirms the cleanliness of the $UO_2$(111) surface.

NO adsorption experiments were performed at a sample temperature of 113 K by cooling the crystal with liquid nitrogen and backfilling the IR chamber using a leak-valve-based directional doser. To protect the UHV system, a gas mixture containing 2000 ppm of NO in $N_2$ was used throughout the experiments. During IRRAS data acquisition, the base pressure was approximately $1\times10^{-10}$ mbar. Prior to each NO exposure, a spectrum of the clean $UO_2(111)$ surface was recorded as a background reference. NO exposures were carried out at an $NO/N_2$ pressure of $1\times10^{-6}$ mbar for 100 s. Further exposure produced no detectable changes in the experimental spectra, confirming the formation of saturated NO adlayer. IRRAS spectra were acquired using p- and s-polarized light at a fixed grazing incidence angle of 80°.

## 2. Computational details and models

First-principles density functional theory (DFT) calculations were performed within the projector augmented-wave (PAW) formalism, as implemented in the Vienna *Ab initio* Simulation Package (VASP) [23-24]. The valence configurations included 14 electrons for U ($6s^2 6p^6 5f^2 7s^2 6d^2$), 6 for O ($2s^2 2p^4$), and 5 for N ($2s^2 2p^3$). Several levels of electronic-structure theory were considered to describe the $NO/UO_2$ system. Because both the localized uranium 5f states and the molecular orbitals of NO are affected by self-interaction errors, approaches beyond local and semilocal density functionals are required. The results were obtained using the screened hybrid Heyd-Scuseria-Ernzerhof functional (HSE06) [15-16] as implemented in VASP, within the non-collinear magnetic formalism.

To assess the influence of magnetic order, both collinear ferromagnetic (CL-FM) and non-collinear (NCL) calculations were performed, following the approach used in Ref. 14. In collinear calculations, the magnetic moments are constrained to lie parallel or antiparallel to a single fixed quantization axis, chosen here as the z axis. In non-collinear calculations, the magnetic moments may point in arbitrary directions, $\mathbf{m}_i = (m_{ix}, m_{iy}, m_{iz})$, and the electronic minimization is performed using the full 2×2 spin-density matrix [25].

Relativistic effects play an important role in the electronic and crystal structure of $UO_2$. Their influence was assessed by performing calculations both with and without spin–orbit coupling (SOC). In VASP, SOC is implemented within the non-collinear formalism [26]. Accordingly, all calculations including SOC were performed non-collinearly. Hereafter, HSE06 (CL), HSE06 (NCL), and HSE06+SOC (NCL) denote HSE06 calculations performed within the collinear formalism, the non-collinear formalism, and the non-collinear formalism including spin–orbit coupling, respectively.

For comparison, PBE+U calculations were also performed using the rotationally invariant Dudarev formulation [27], with an effective Hubbard parameter $U_{eff}$= 4 eV applied to the uranium 5f states. The choice of this value is discussed in Ref. 14. The relaxed equilibrium lattice parameters of bulk $UO_2$ were calculated specifically for the collinear PBE+U, collinear HSE06, and non-collinear HSE06+SOC approaches (Table 1 of Ref. 14).

The $UO_2$(111) surface was modelled using 1 × 1, 2 × 1, 3 × 1, and 2 × 2 surface unit cells consisting of three O–U–O trilayers stacked along the surface normal (z-axis). A vacuum region of approximately 15 Å was introduced along the surface-normal direction to prevent spurious interactions between periodically repeated slabs. All calculations were performed using a plane-wave kinetic-energy cutoff of 450 eV. Brillouin-zone integrations were carried out using Γ-centered Monkhorst-Pack k-point meshes of 3×3×1 grid for the 1×1 surface cell, 1×3×1 for the 2×1 and 3×1 cells, and 1×1×1 grid for the larger 2×2 surface cell.

During geometry optimization, the two top trilayers were fully relaxed, whereas the bottom trilayer was kept fixed at bulk positions. Structural relaxation was continued until the residual forces on all relaxed atoms were below 0.01 eV/Å.

The search for stable adsorption sites was performed by placing the NO molecule at the 9 different initial positions shown in Fig. S3, followed by full structural relaxation from each starting configuration. During these relaxations, the two outermost $UO_2$ trilayers and the adsorbed NO molecule were allowed to relax according to the same convergence criterion described above. Owing to the absence of inversion symmetry in the slabs, dipole corrections, as implemented in VASP, were included in all calculations. In addition, dispersion interactions were described using Grimme's D3 correction with Becke–Johnson damping [18] to obtain a more accurate description of NO adsorption on the $UO_2$ surface. Among the nine initial configurations considered, only two converged to stable adsorption geometries, which are discussed in the main text (Table I and Figs. 1(c) and 1(d)).

Charge-density-difference plots were generated using the VESTA [28] and VASPKIT [29] packages.

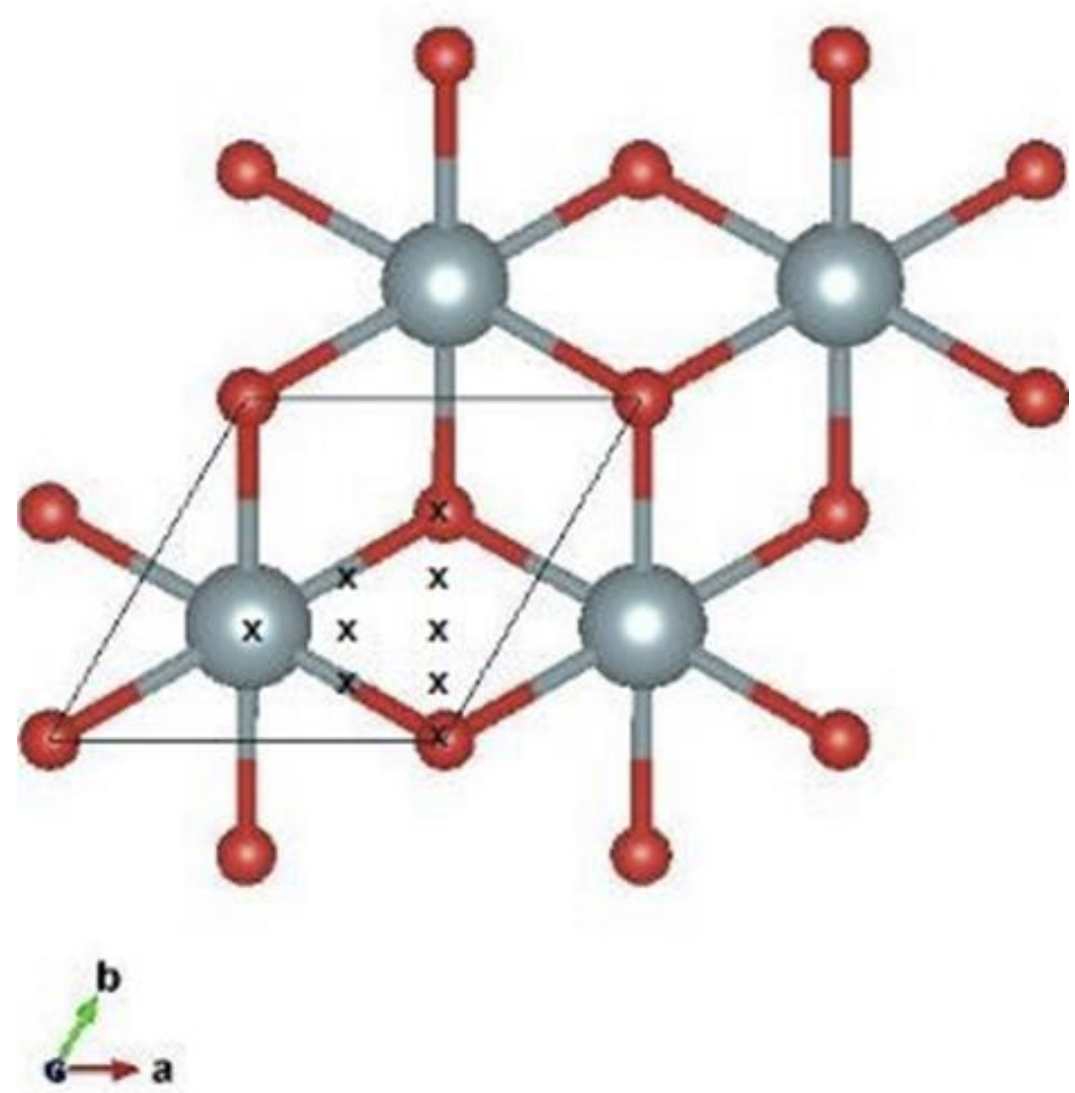

**Figure S3:** 9 initial positions (crosses) of the NO molecule considered in the search for stable adsorption sites on the 1×1 $UO_2(111)$ surface.

The vibrational frequencies of the different adsorption geometries were calculated within the harmonic approximation using the finite-difference method to evaluate the second derivatives of the total energy with respect to the atomic displacements. The force-constant matrix was constructed by displacing only the atoms of the outermost $UO_2$ trilayer and the NO molecule, while all remaining atoms were kept fixed.

The infrared intensity associated with each normal mode is directly proportional to the square of the first derivative of the surface-normal component of the dipole moment with respect to the corresponding normal coordinate, $(\partial \mu z/\partial Q)^2$.

Accordingly, the heights of the calculated peaks shown in Fig. 1a are proportional to the normalized IR intensities, which were first normalized per adsorbed NO molecule and scaled such that the maximum intensity equals unity. In Table 1, the superscripts *s, m,* and *w* denote qualitatively strong, medium, and weak normalized IR intensities, respectively. Specifically, modes with normalized intensities greater than 0.05 are labeled *s*, those between 0.01 and 0.05 are labeled *m*, and those between 0.001 and 0.01 are labeled *w*.

## 3. Collective modes in larger supercells

To investigate the collective vibrational modes of the NO adlayer, 1×2, 1×3, and 2×2 supercells with one monolayer of adsorbed NO were considered. The relaxed adsorption structures together with the corresponding collective vibrational modes involving the NO molecules are shown in Fig. S4. Animated visualizations of collective modes for structure 5 (Supplementary Movie S1 and Supplementary Movie S2), structure 9 (Supplementary Movie S3 and Supplementary Movie S4), and structure 11 (Supplementary Movie S5 and Supplementary Movie S6) of Fig. S4 are provided as Supplementary Movies S1–S6.

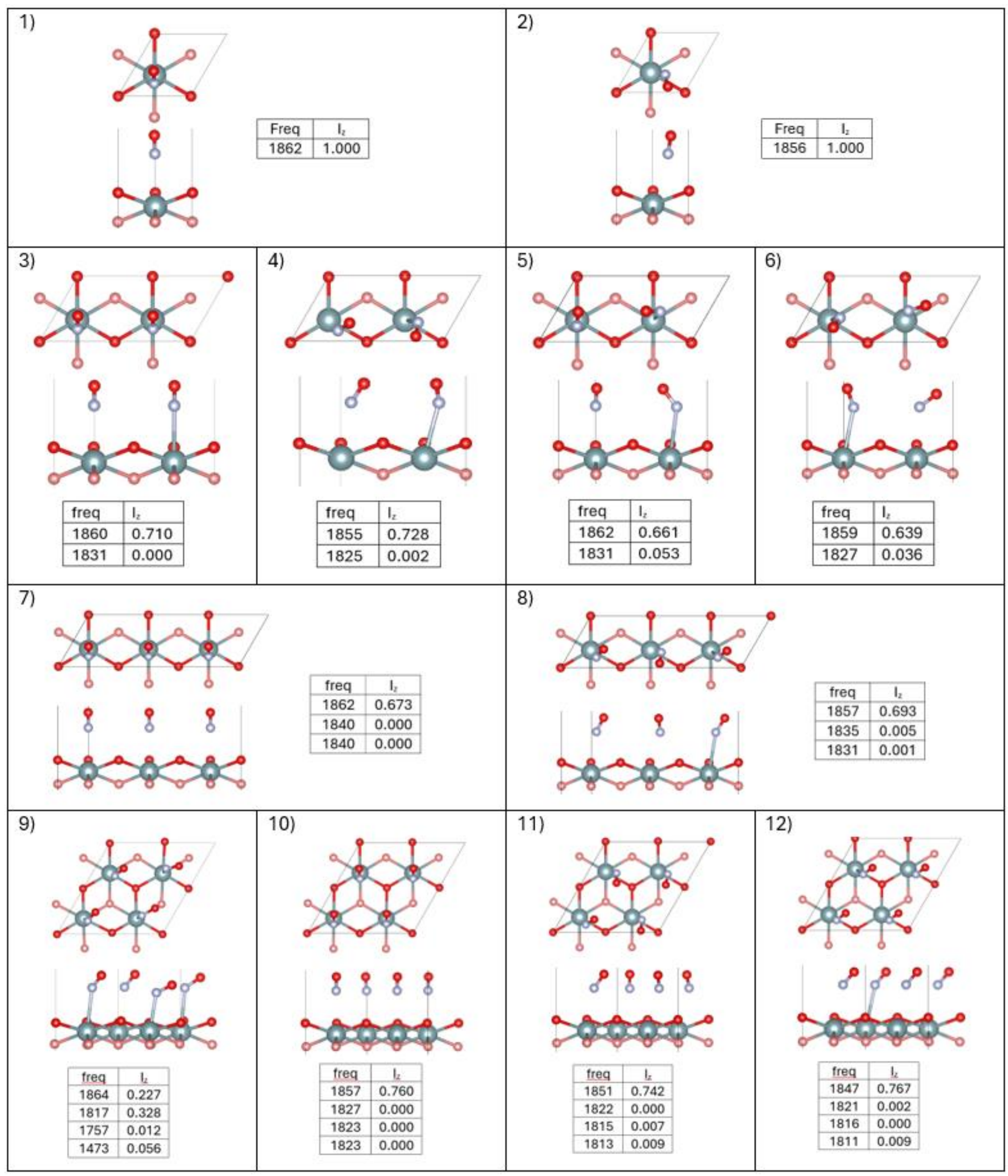

**Figure S4:** Top and side views of each of the relaxed 1×2, 1×3, and 2×2 $UO_2$(111) supercells with one monolayer of adsorbed NO, together with the corresponding collective vibrational modes involving the NO molecules.

## 4. Effect of the level of theory on the adsorption properties of NO on $UO_2$(111)

Before discussing the adsorption properties of NO on $UO_2$(111), it is instructive to examine how the different levels of electronic-structure theory describe the isolated NO and CO molecules. To the best of our knowledge, no experimental value for the gas-phase HOMO–LUMO gap of NO has been reported. For CO, the calculated HOMO–LUMO gaps are 6.8 eV and 9.2 eV at the PBE and HSE06 levels, respectively. In contrast, the corresponding values for NO are only 0.2 eV and 2.8 eV. Figure S5 presents the charge-density difference and projected density of states (PDOS) for NO on $UO_2$(111) calculated using the PBE+U approach. The extremely small HOMO–LUMO gap predicted for NO at the PBE level indicates that this functional substantially overestimates the electronic softness of the molecule and is therefore expected to overestimate its interaction with the substrate.

**Table S1**: N–O bond length, stretching frequency, and scaling factor, $\lambda=\nu_{exp,gas}/\nu_{calc,gas}$, with $\nu_{exp,gas}$ = 1886 $cm^{-1}$, for the isolated NO molecule calculated in a 10 × 11 × 12 $Å^3$ supercell.

| **Electronic-structure treatment** | **$d_{NO}$ [pm]** | **$\nu$ [$cm^{-1}$]** | **$\lambda=\nu_{exp,gas}/\nu_{calc,gas}$** |
|---|---|---|---|
| **PBE+U (CL)** | 117.01 | 1924 | 0.974976 |
| **HSE06 (CL)** | 115.55 | 2066 | 0.908101 |
| **HSE06+SOC (NCL)** | 115.55 | 2066 | 0.907995 |

**Table S2:** Structural, magnetic, and vibrational properties of NO adsorbed on $UO_2(111)$ calculated using different levels of electronic-structure descriptions. The table reports the magnetic moments of the adsorbed NO molecule and of the uranium atoms, the N–O bond length $d_{NO}$, its variation with respect to the gas-phase molecule, $\Delta d_{NO} = d_{NO} - d_{NO,gas}$, the N–U bond distance, $d_{NU}$, the scaled N–O stretching frequency, $\nu = \lambda\, \nu_{calc}$, and the corresponding frequency shift with respect to the scaled gas-phase frequency, $\Delta\nu = \nu - \nu_{gas}$, where $\nu_{gas} = \lambda\, \nu_{calc,gas}$. The scaling factor λ is given in Table S1.

| **Electronic-structure treatment** | **Magnetic moments ($\mu_B$)** | **$d_{NO}$ [pm]** | **$\Delta d_{NO}$ [pm]** | **$d_{NU}$ [Å]** | **$\nu$ [cm$^{-1}$] (scaled)** | **$\Delta\nu$ [cm$^{-1}$]** |
|---|---|---|---|---|---|---|
| **PBE+U (CL)** | $m_{U1}$=( 0.00, 0.00, 2.05)<br>$m_{U2}$=( 0.00, 0.00, 2.03)<br>$m_{U3}$=( 0.00, 0.00, 2.03)<br>$m_{NO}$=( 0.00, 0.00, 0.78) | 117.56 | +0.55 | 2.67 | 1818 | -58 |
| **HSE06 (CL)** | $m_{U1}$=( 0.00, 0.00, 1.98)<br>$m_{U2}$=( 0.00, 0.00, 1.98)<br>$m_{U3}$=( 0.00, 0.00, 1.52)<br>$m_{NO}$=( 0.00, 0.00, 1.00) | 117.81 | +2.26 | 2.23 | 1712 | −164 |
| **HSE06 (NCL, no SOC)** | $m_{U1}$=( 0.05, 0.04, 1.98)<br>$m_{U2}$=( 1.47, 0.29, 1.28)<br>$m_{U3}$=( 1.65, 0.03, 0.14)<br>$m_{NO}$=(-0.49,-0.01,-0.04) | 116.50 | +0.95 | 2.39 | 1788 | -87 |
| **HSE06+SOC (NCL), SAXIS = (001)**<br>$E_{tot}$= -143.138 eV | $m_{U1}$=(-0.05, 0.09, 1.47)<br>$m_{U2}$=( 0.56,-1.04,-0.95)<br>$m_{U3}$=( 1.40, 0.00, 0.48)<br>$m_{NO}$=(-0.66, 0.11,-0.21) | 115.54 | -0.01 | 2.82 | 1862 | -14 |
| **HSE06+SOC (NCL), SAXIS = (110)**<br>$E_{tot}$ = -143.196 eV | $m_{U1}$=( 0.45, 0.42, 1.35)<br>$m_{U2}$=( 0.91,-1.15,-0.40)<br>$m_{U3}$=( 0.29, 0.43, 1.36)<br>$m_{NO}$=( 0.64,-0.12,-0.27) | 115.53 | -0.02 | 2.81 | 1862 | -14 |
| **HSE06+SOC (NCL), SAXIS = (001)**<br>$E_{tot}$ = -143.138 eV | $m_{U1}$=(-0.02, 0.00, 1.48)<br>$m_{U2}$=( 1.18, 0.00, 0.98)<br>$m_{U3}$=( 1.40,-0.07, 0.48)<br>$m_{NO}$=(-0.66, 0.13, 0.19) | 115.54 | -0.01 | 2.82 | 1862 | -14 |

The total energies, $E_{tot}$, indicated only for the SOC calculations of Table S2, shows the relative stability of the different magnetic solutions (see also section 6). Note that two of them present the same $E_{tot}$ and quantization axis but different magnetic configurations, evidencing the complex ground state of this system.

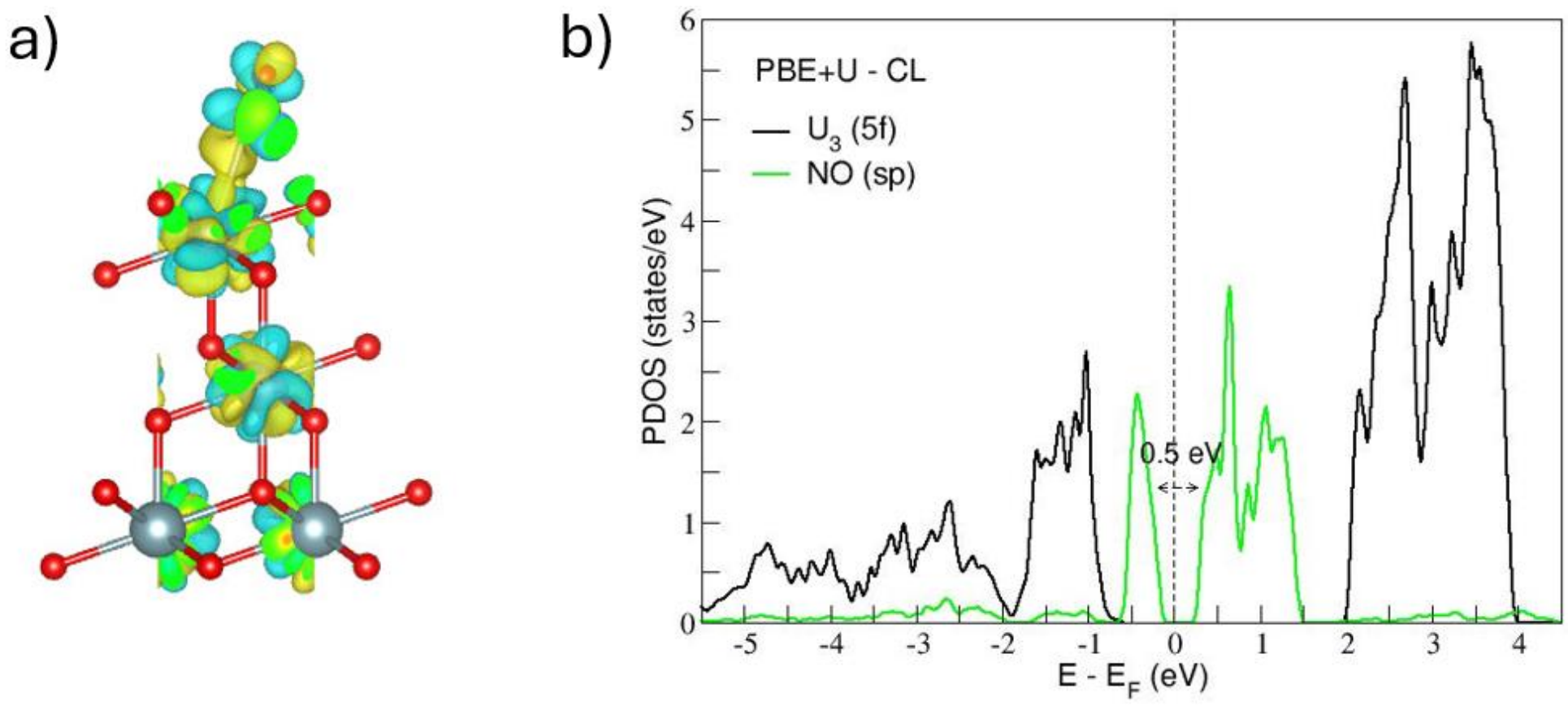

**Figure S5:** Charge-density difference (a) and projected density of states (PDOS) (b) for NO adsorbed on $UO_2$(111), calculated using the PBE+U for $UO_2$ and PBE for the NO molecule. The PDOS correspond to the 5f states of the surface uranium atom directly bonded to NO ($U_3$) and to the molecular orbitals of the adsorbed NO molecule. The charge-density difference is shown using an isosurface value of ±0.0034 e $Å^{-3}$; yellow and light-blue isosurfaces denote charge accumulation and depletion, respectively.

To further assess the role of dispersion interactions, additional calculations were performed for configurations 1 and 9 of Fig. S4 without van der Waals corrections. For configuration 1, omitting dispersion causes the NO molecule to relax farther from the surface, increasing the N–U distance to $d_{NU}$ = 2.88 Å. In contrast, the internal N–O bond length and the stretching frequency remain essentially unchanged, with $d_{NO}$ = 115.52 pm and $\nu$= 1865 $cm^{-1}$, respectively.

For configuration 9, omitting the van der Waals correction significantly weakens the lateral interactions between neighboring NO molecules (Fig. S6). As a consequence, the distribution of oscillator strength among the collective vibrational modes is modified, with the fully symmetric in-phase stretching mode gaining relative intensity, while the remaining collective vibrational modes become weaker. These calculations further support the conclusion that dispersion-driven intermolecular interactions play a key role in the emergence of the manifold of collective vibrational modes responsible for the broad asymmetric infrared band discussed in the main text.

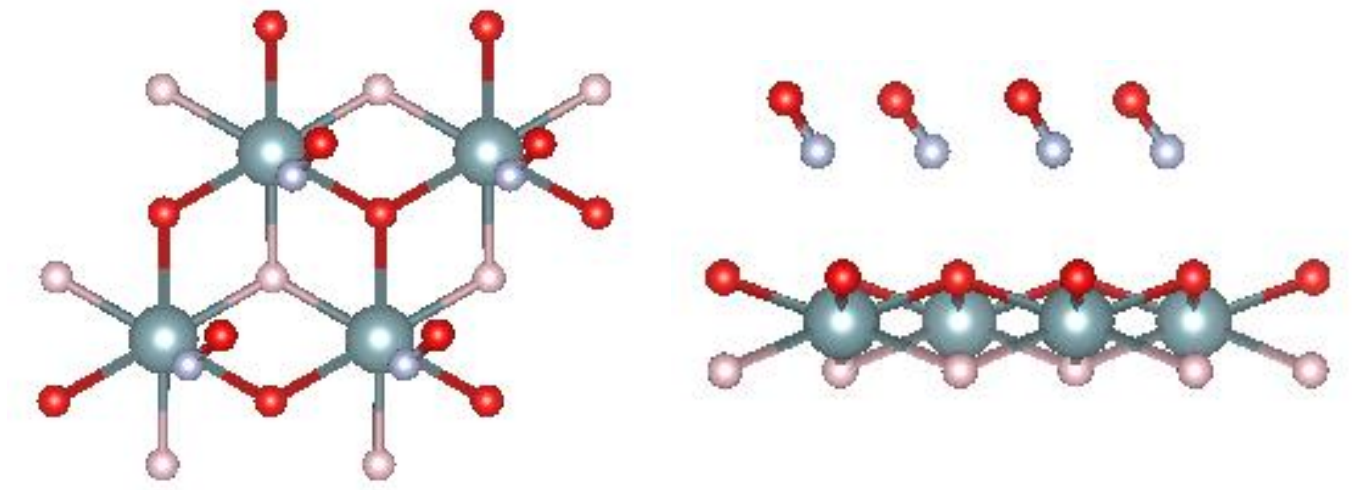

| freq | $I_z$ |
|---|---|
| 1858 | 0.716 |
| 1834 | 0.002 |
| 1828 | 0.000 |
| 1825 | 0.002 |

**Figure S6:** Top and side views of configuration 9 shown in Fig. S4, calculated without van der Waals corrections, together with the corresponding collective vibrational modes involving the NO molecules. Intensities are scaled as in Table S2.

## 5. Simple Mechanical Model for the Vibrational Shift

A qualitative interpretation of the N–O stretching vibrational response can be obtained by considering the interplay between the intrinsic N–O bond stiffness and the restoring force associated with the adsorbate–substrate interaction. This effect can be illustrated using a simple two-spring model, in which one spring represents the internal N–O bond ($K_{NO}$) and the other the adsorbate–substrate interaction ($K_{NO_sub}$). In this simplified picture, the two springs act in series, yielding an effective force constant

$$K_{eff} = K_{NO} * K_{NO_sub} /(K_{NO} + K_{NO_sub}).$$

For example, if $K_{NO_sub} = 0.1\ K_{NO}$, then $K_{eff} \approx 0.09\ K_{NO}$. Although this simple model is intended only as a qualitative illustration, it shows how coupling to the substrate can reduce the effective restoring force of the N–O stretching mode, leading to a red shift even when the equilibrium N–O bond length changes only slightly.

## 6. Influence of the magnetic structure of $UO_2$ on NO adsorption

To evaluate the sensitivity of the adsorption energetics to the magnetic structure of the $UO_2$ substrate, we also considered several non-collinear magnetic configurations of the slab for each adsorption geometry discussed in the main text. Figure S7 summarizes the corresponding total energies relative to the lowest-energy magnetic configuration for each adsorption structure. The energy spread of energies associated with the different magnetic configurations is typically within ±35 meV (corresponding to a maximum difference of ~70 meV), providing an estimate of the effect of the magnetic ordering of the $UO_2$ slab on the adsorption energy. As discussed in the main text, these variations have a negligible influence on the calculated N–O stretching frequencies.

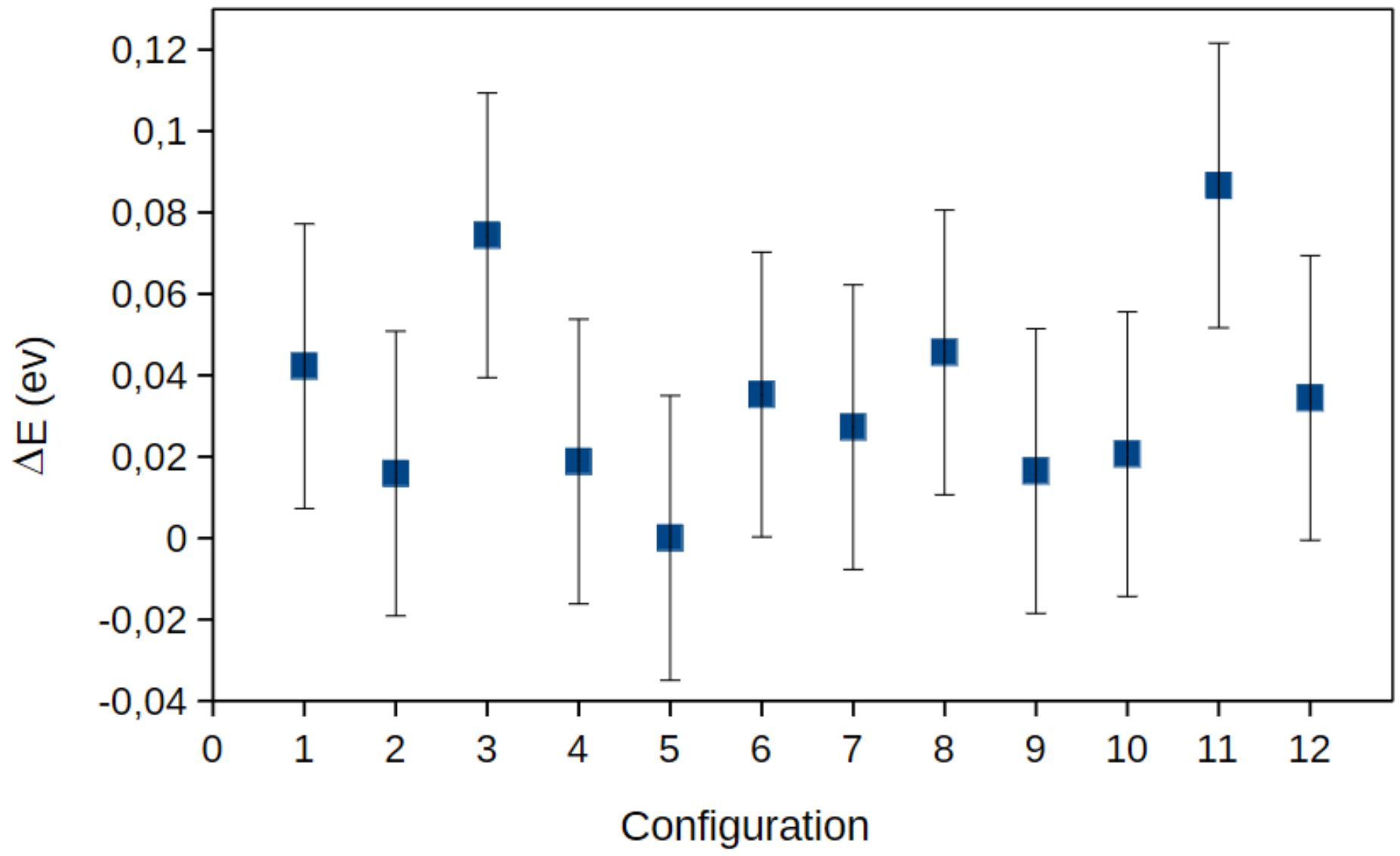

**Figure S7:** Total energy differences (per unit cell) with respect to the lowest-energy magnetic configuration for each adsorption structure shown in Fig. S4. The error bars (±35meV) represent the energy spread associated with different magnetic configurations of the $UO_2$ slab.

## 7. Magnetic coupling between the molecule and the $UO_2$

To assess the influence of the magnetic coupling between the magnetic moments of the NO molecule and the surface uranium atom ($U_3$), we performed constrained non-collinear calculations for several fixed orientations of the NO magnetic moment relative to that of $U_3$. These calculations were carried out using the "I_CONSTRAINED_M = 4" scheme implemented in VASP, which constrains both the direction and the sign of the magnetic moments [30]. In our calculations, the penalty parameter λ, which enters the constraint energy functional, was set to 50, resulting in a penalty energy $E_P$ below $1\times10^{-4}$ eV. Figure S8 shows the energy differences between selected constrained magnetic configurations. The magnetic moment of the $U_3$ atom (blue arrow) was fixed to $\boldsymbol{m}_{U3}$=(1.1,0.0,1.0) $\mu_B$, corresponding to a reference magnetic state, while the orientation of the NO magnetic moment was systematically varied (green arrows). The total energies obtained for the different relative orientations span less than 20 meV, indicating that the magnetic coupling between the NO molecule and the surface uranium atom is weak

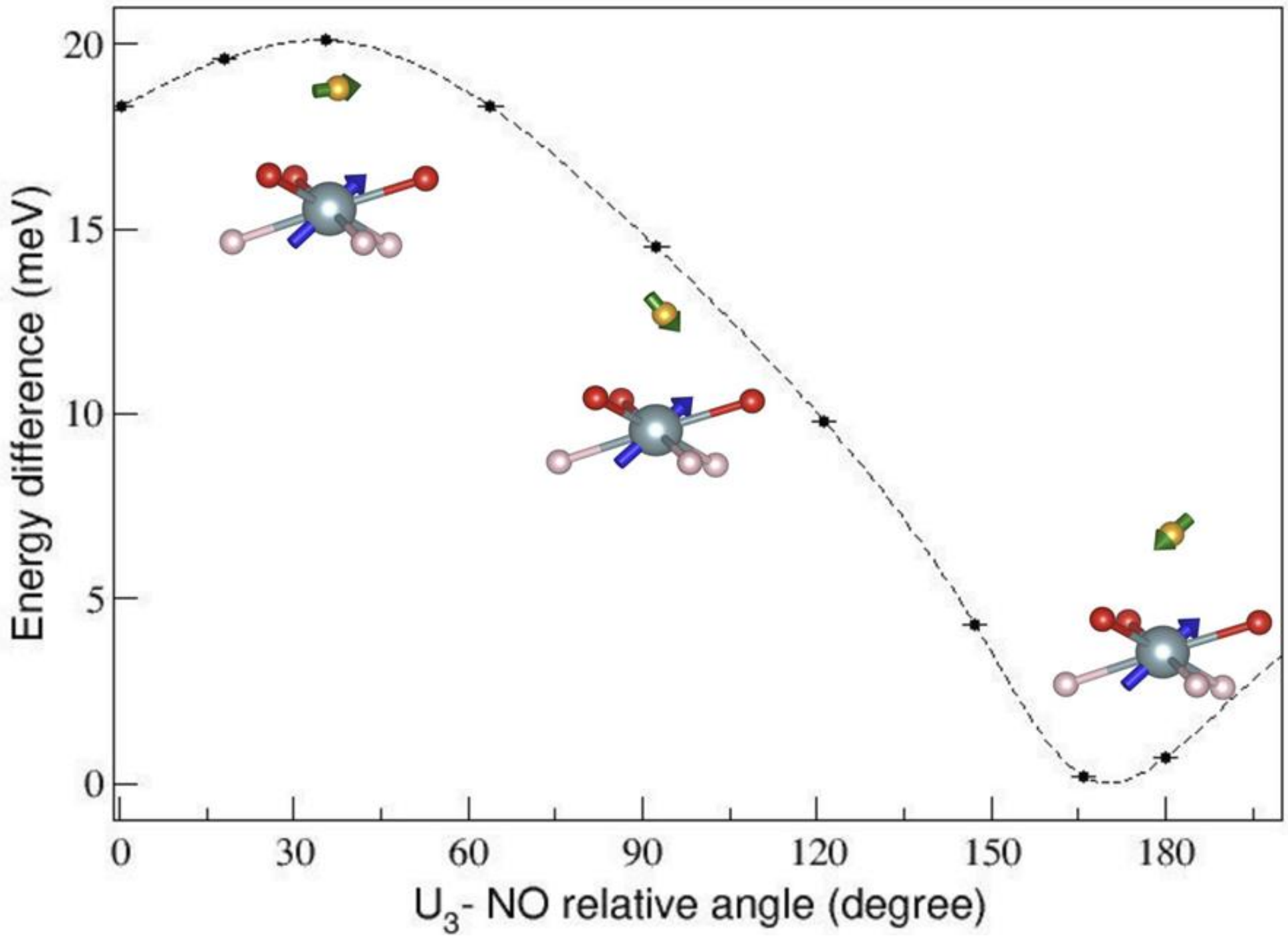

**Figure S8:** Energy (meV) as a function of the relative angle (deg) between the $U_3$ and NO molecule magnetic moments obtained from constrained noncollinear calculations. The blue and green arrows represent the $U_3$ and NO magnetic moments, respectively. The blue arrow is fixed, whereas the green arrows illustrate selected orientations of the NO magnetic moment relative to that of $U_3$ (30°, 90°, and 180°). The error bars are proportional to the penalty energy, $E_P$. The dashed line is included as a guide to the eye.

## 8. Intermolecular Interaction between NO molecules

To further investigate the sources of the intermolecular interaction, we performed calculations for pairs of NO and CO molecules in vacuum while systematically varying the intermolecular separation. During the structural relaxations, the positions of the N atoms were kept fixed for NO and, analogously, the positions of the C atoms were fixed for CO. Figure 3 shows the total energy of the NO dimer as a function of the intermolecular distance, d. The energies are referenced to the configuration with the largest separation (d=10 Å). As expected, the interaction is attractive at intermediate intermolecular separations. At shorter distances, overlap of the molecular charge densities leads to the onset of short-range Pauli repulsion, making the interaction progressively less attractive.